\documentclass[a4paper,10pt]{article}

\usepackage[pdftex]{color,graphicx,hyperref}
\usepackage{amsmath,amsfonts,amssymb}
\usepackage{fullpage}
\usepackage{enumerate}
\usepackage{sidecap}
\usepackage{lscape}
\usepackage{authblk}
\usepackage{appendix}
\usepackage{setspace}
\usepackage[numbers,sort&compress]{natbib}
\usepackage[labelfont=bf,labelsep=period]{caption}

\def\kvec{\mathbf{k}}
\def\Pk{p_{\kvec}}
\def\Fkm{F_{\kvec, m}}
\def\skm{s_{\kvec, m}}
\def\skmo{s_{\kvec, m-1}}
\def\dskm{\dot{s}_{\kvec, m}}
\def\pk{\rho_{\kvec}}
\def\sumk{\sum_{\kvec}}
\def\summ{\sum_m}
\def\summLess{\sum_{m < k\phi}}
\def\summMore{\sum_{m \geq k\phi}}
\def\bs{\beta_s}
\def\Bkm{B_{k, m}}
\def\Bkom{B_{k-1, m}}
\def\Bkmo{B_{k, m-1}}
\def\ft{f_t}

\title{Kinetics of social contagion}

\author[1,2]{Zhongyuan Ruan}
\author[3,4]{Gerardo I\~{n}iguez}
\author[5]{M\'{a}rton Karsai}
\author[1,2,4]{J\'{a}nos Kert\'{e}sz \thanks{janos.kertesz@gmail.com}}

\affil[1]{\small{Center for Network Science, Central European University, H-1051 Budapest, Hungary}}
\affil[2]{Institute of Physics, Budapest University of Technology and Economics, H-1111 Budapest, Hungary}
\affil[3]{Centro de Investigaci{\'o}n y Docencia Econ{\'o}micas, Consejo Nacional de Ciencia y Tecnolog{\'\i}a, 01210 M{\'e}xico D.F., Mexico}
\affil[4]{Department of Computer Science, Aalto University School of Science, FI-00076 AALTO, Finland}
\affil[5]{Laboratoire de l'Informatique du Parall\'elisme, INRIA-UMR 5668, IXXI,  ENS de Lyon, 69364 Lyon, France}

\begin{document}

\maketitle

\begin{abstract}

Diffusion of information, behavioral patterns or innovations follows diverse pathways depending on a number of conditions, including the structure of the underlying social network, the sensitivity to peer pressure and the influence of media. Here we study analytically and by simulations a general model that incorporates threshold mechanism capturing sensitivity to peer pressure, the effect of `immune' nodes who never adopt, and a perpetual flow of external information. While any constant, non-zero rate of dynamically-introduced spontaneous adopters leads to global spreading, the kinetics by which the asymptotic state is approached shows rich behavior. In particular we find that, as a function of the immune node density, there is a transition from fast to slow spreading governed by entirely different mechanisms. This transition happens below the percolation threshold of network fragmentation, and has its origin in the competition between cascading behavior induced by adopters and blocking due to immune nodes. This change is accompanied by a percolation transition of the induced clusters.

\end{abstract}

There are remarkable analogies between the social contagion of information, behavioral patterns or innovation and some physical or epidemic spreading processes, where global phenomena emerge through the diffusion of microscopic states \cite{barrat2008dynamical,easley2010networks,goffman1964generalization,jackson2008social}. All evolve in networks with nodes characterized by relevant state variables, and links that represent direct interactions between nodes. In biological systems epidemics are driven by binary interactions that lead to the emergence of {\it simple contagion} phenomena \cite{barrat2008dynamical}. Social diffusion processes are usually characterized by {\it complex contagion} mechanisms, where node states are determined by comparing individual thresholds with all neighbor states \cite{easley2010networks,karsai2014complex,watts2002simple,wejnert2002integrating, centola2007complex}. This property, capturing the effect of peer pressure and commonly assumed in social spreading phenomena \cite{granovetter1983threshold,centola2010spread}, has consequences on the dynamics and the final outcome of the social contagion process. Moreover, the theoretical approach to these systems has much in common \cite{barrat2008dynamical,watts2002simple,gleeson2007seed}, which greatly helps us to understand their behavior.

Models employing threshold mechanisms mostly focus on cascading phenomena where, under some circumstances, a macroscopic fraction of nodes in the network is converted rapidly due to microscopic perturbations. This approach is motivated by earlier social theories \cite{granovetter1983threshold, schelling1969models} and has been implemented by Watts in an elegant model of cascading behavior \cite{watts2002simple}. Watts showed that a global cascade (occupying a macroscopic fraction of the network and induced by local perturbations) can occur due to the interplay between network structure and individual thresholds. He further identified the phase with a non-zero probability of global cascades in the space $(\phi, z)$ of the average threshold $\phi$ of nodes and the average degree $z$ of the network.

\begin{figure}
\centering
\includegraphics[width=0.8\linewidth]{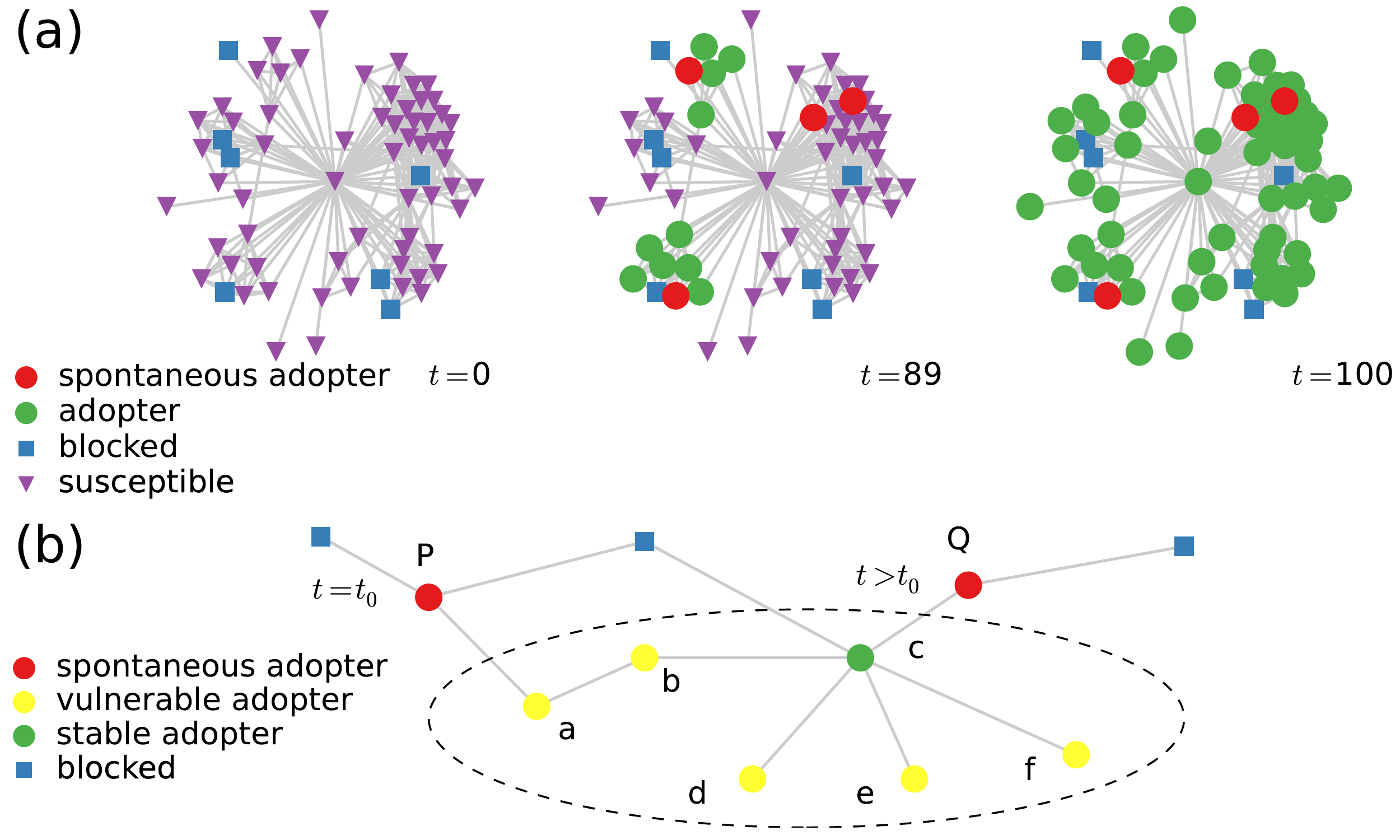}
\caption{(color online). (a) Numerical simulation of a general threshold model over an empirical network, with adoption threshold $\phi = 0.2$, rate of spontaneous adopters $p = 0.0005$ and fraction of blocked nodes $r = 0.1$. The network is an ego sample of Facebook friendships with size $N = 96$ and average degree $z = 10.63$ \cite{leskovec2012learning}. Susceptible nodes adopt spontaneously with rate $p$ or after a fraction $\phi$ of their neighbors has adopted, while blocked nodes never adopt. (b) Schematic illustration of the spreading process. At $t = t_0$ node $\mathrm{P}$ spontaneously becomes an adopter, `infecting' nodes $\mathrm{a}$ and $\mathrm{b}$. When $\mathrm{Q}$ adopts, it induces the adoption of nodes $\mathrm{c}-\mathrm{f}$. Nodes inside the ellipse constitute an induced cluster of adoption.}
\label{Fig:1}
\end{figure}

While the relevance of this model is indisputable \cite{watts2002simple,gleeson2007seed,watts2007influentials,gleeson2013binary,gleeson2011high,gleeson2008cascades,centola2007cascade,payne2009information,yaugan2012analysis,nematzadeh2014optimal,singh2013threshold,piedrahita2013modeling},
its limitations become clear from real social spreading data. The Watts model focuses on the (instantaneous) emergence of global cascades triggered by single local perturbations, while there are empirical examples where threshold mechanisms do play a role yet global adoption phenomena emerge through other scenarios. In reality global adoption is often not induced by microscopic perturbations but by a larger fraction of people \cite{singh2013threshold}. Moreover, decisions of individuals depend on external impulses arriving from mass media or advertizing \cite{kocsis2011competition}, resulting in a perpetual stochastic perturbation. In addition, there are individuals entirely reluctant to adopt. Furthermore, the Watts criterion for macroscopic adoption is purely deterministic, coded in the network structure, threshold distribution and perturbation site -- it does not concern time, which is clearly a feature of empirical stochastic processes of adoption spreading.

Here we present a general threshold-driven model of social contagion phenomena that captures various spreading scenarios, ranging from cascading behavior to dynamically evolving non-explosive patterns, and sheds light to the different kinetics behind them (Fig.~\ref{Fig:1}). Motivated by empirical observations \cite{karsai2015anatomy}, we extend Watts' threshold model by considering {\it blocked} nodes immune to social influence and discuss their effect on cascade formation. In addition, we introduce spontaneous adopters with a constant rate, and present approximate analytical and numerical results regarding our model. In particular, we study how the kinetics of spreading changes for an increasing density of blocked nodes. We aim at the simplest possible but sufficiently general extension of earlier threshold models \cite{watts2002simple,centola2007cascade,payne2009information,yaugan2012analysis,nematzadeh2014optimal,singh2013threshold,piedrahita2013modeling} with a minimal set of states and transitions necessary to describe various real scenarios of social spreading phenomena. The introduction of further states, secondary adoption, or other decision-making mechanisms is left as a further challenge, since our aim here is to model generic cascades of primary adoption.

\begin{figure}
\centering
\includegraphics[width=0.8\linewidth]{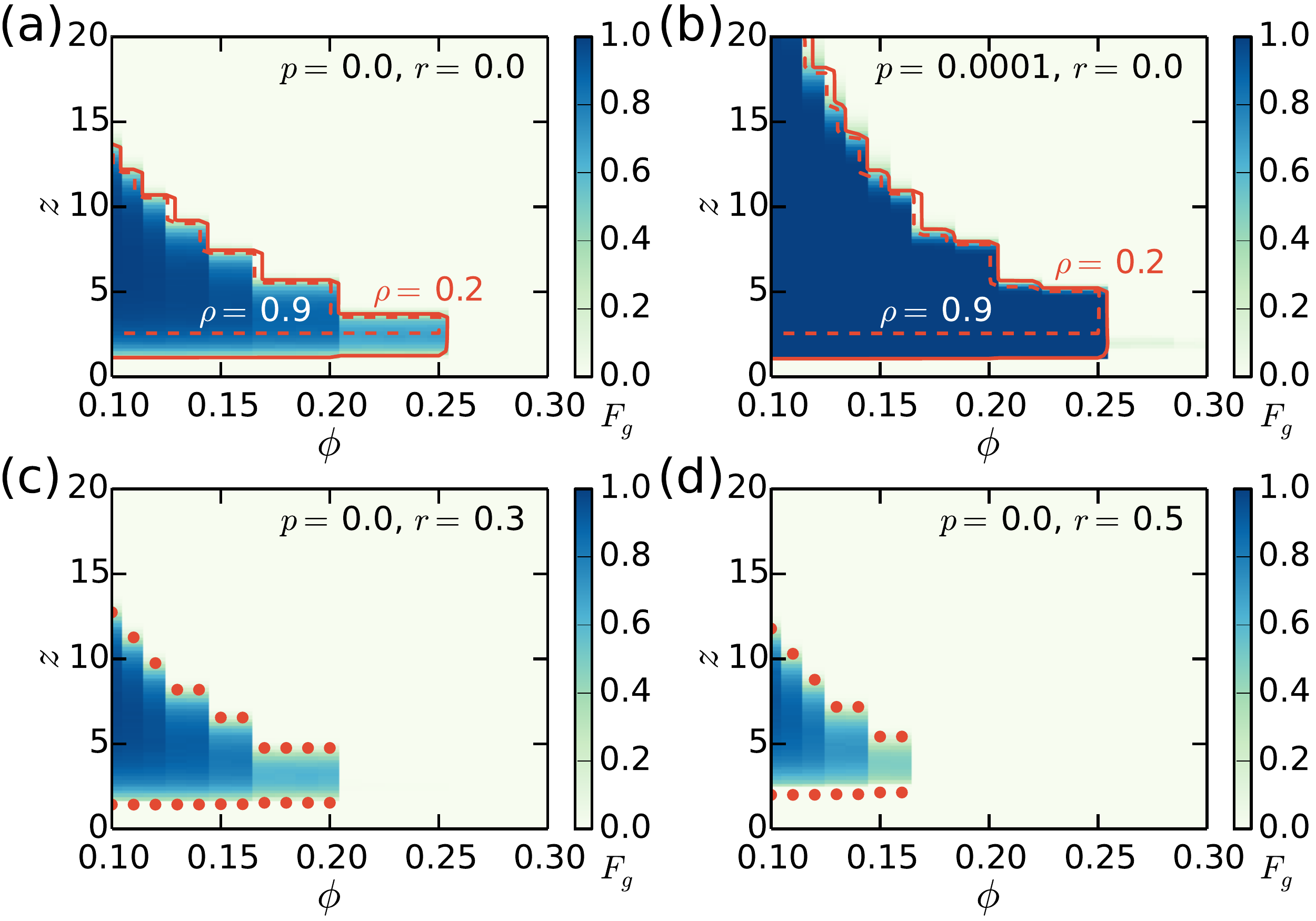}
\caption{(color online). Frequency $F_g$ of global cascades as a function of node threshold $\phi$ and network average degree $z$ at an intermediate time $t = 100$, for varying $p$ and $r$. (a-b) As $p$ increases, the region that allows global cascades of adoption grows in size. Its boundary is well approximated by a cut-off in the fraction of adopters $\rho$ as calculated by Eq.~(\ref{eq:reducedAMEs}). (c-d) Conversely, an increasing fraction of blocked nodes shrinks the global cascade regime. Dots show the boundary of this regime according to Eq.~(\ref{eq:critical2}). Simulations correspond to an ER network with $N=10^4$ and are averaged over $10^4$ realisations.}
\label{Fig:2}
\end{figure}

In Watts' threshold model \cite{watts2002simple}, all nodes are initially in a susceptible state $0$, except for a single adopter seed in state $1$. The process evolves as each node with degree $k$ changes its state from $0$ to $1$ if a fraction $\phi$ of its neighbors have adopted before. Since nodes cannot change their state after exposure, the system evolves towards a state where no further adoptions are possible. The emergence of a global cascade depends on the degree distribution $p_k$ of the network, the distribution $p_{\phi}$ of individual thresholds, and the initial seed. The condition for a global cascade is the existence of a percolating component of {\it vulnerable} nodes with thresholds $0 < \phi \leq 1/k$ (who need one adopting neighbor before exposure) connected to the seed. This percolating vulnerable tree is quickly converted after adoption of the seed and may trigger further adoption of {\it stable} nodes with thresholds $\phi > 1/k$ (who need more than one adopting neighbor to adopt). Assuming an Erd\H{o}s-R\'{e}nyi (ER) random network \cite{erdos1960evolution} and a single adopter seed, there is a phase boundary in $(\phi, z)$-space encompassing a regime where global cascades occur (Fig.~\ref{Fig:2} a). The properties of this cascading regime have been investigated for the case of heterogeneous thresholds, different network topologies \cite{watts2002simple,gleeson2007seed}, and variable seed size \cite{gleeson2008cascades,singh2013threshold}.

Empirical studies, however, support the intuition that some individuals in society may refuse to adopt technological innovations for various reasons -- due to another favorite product, aversion towards a firm, or some criticism on principle~\cite{karsai2015anatomy}. Such individuals will never be exposed, irrespective of the state of their neighbors \cite{yildiz2013binary}. To consider this behavioral pattern in our model, we block the adoption of a fraction $r$ of randomly selected nodes in the network. These nodes do count when their neighbors consider the decision to adopt, and thus will make it harder for neighbors to fulfil the threshold criterion.

Accordingly, the original Watts model corresponds to $r=0$, while for $r > 0$ the phase diagram changes. Even in the presence of blocked nodes, macroscopic spreading is still determined by the static criterion of the existence of a global vulnerable cluster, and thus a generating function technique \cite{watts2002simple} can be applied~\footnote{see Supplemental Material (SM) for details on the analytical treatment of the model.} \cite{newman2001random}. Assuming a single threshold $\phi$ and an ER network with average degree $z$, the condition for the emergence of a macroscopic cascade is,
\begin{eqnarray}
\label{eq:critical2}
(1-r)e^{-z}\sum_{k=2}^{k_c}\frac{z^{k}}{(k-2)!}-z=0,
\end{eqnarray}
with $k_c=\lfloor 1/\phi \rfloor$. Due to the factor $1 - r$, the introduction of blocked nodes shrinks the region in $(\phi, z)$-space where global cascades develop, in good agreement with numerical simulations (Fig.~\ref{Fig:2} c and d).

While blocked nodes hinder the spreading process, there are reasons other than social influence that could motivate individuals to adopt a social pattern, like external influence from mass media. This {\it spontaneous} adoption has been studied theoretically by introducing a given density of adopters at the outset of the Watts model~\cite{singh2013threshold}. However, spontaneous adopters may get active at any time during a real social contagion. Thus we include a stochastic dynamics where a susceptible node may become adopter with rate $p$ at any time, irrespective of the status of its neighbors.

\begin{figure}
\centering
\includegraphics[width=0.7\linewidth]{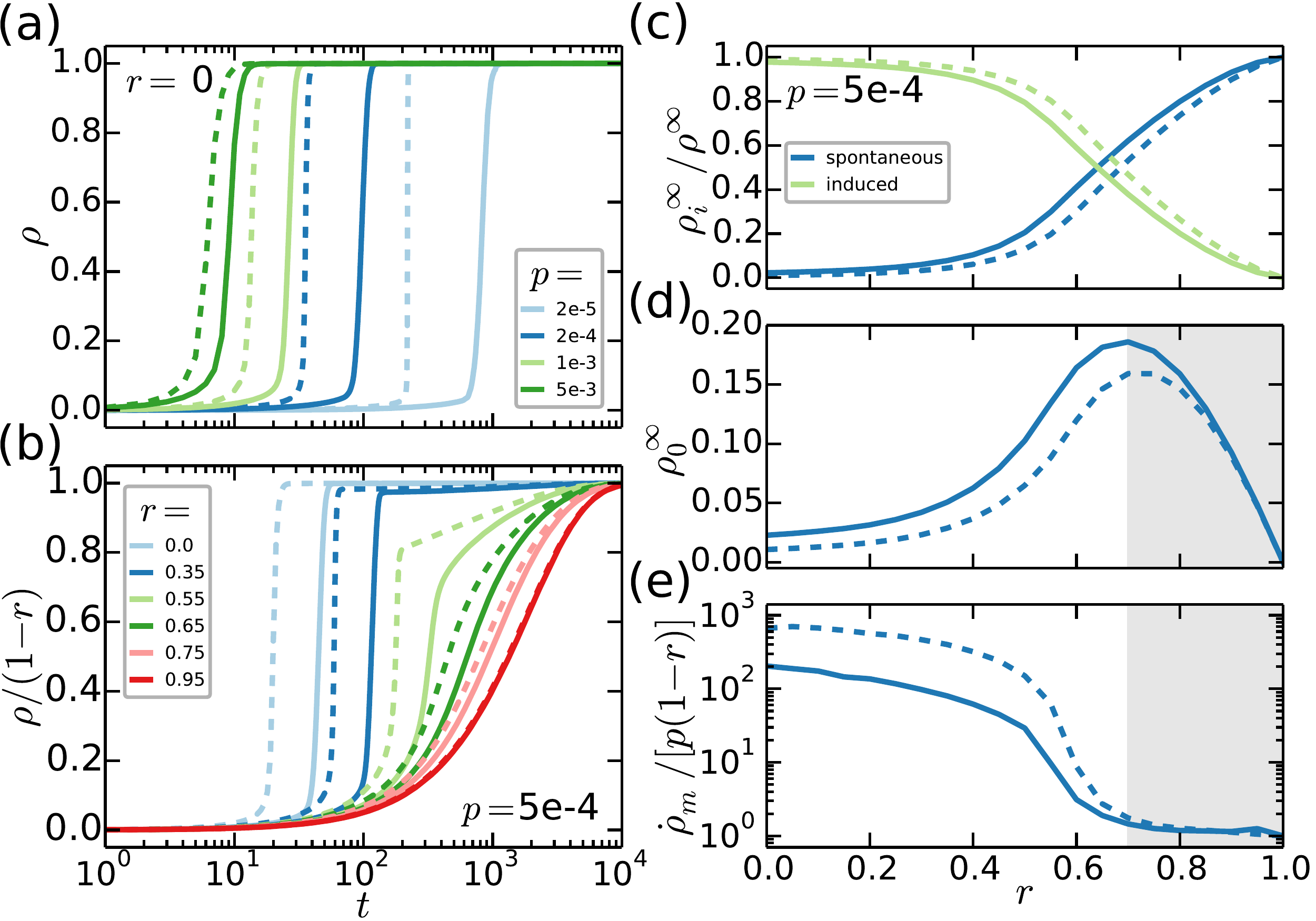}
\caption{(color online). Numerical simulations and analytical approximation of the threshold model for $z=7$ and $\phi=0.2$ (continuous and dotted lines, respectively). (a) Fraction of adopters $\rho$ as a function of time for varying $p$ and fixed $r$. (b) Time evolution of the normalized adoption density $\rho/(1-r)$ for different values of $r$ and fixed $p$. (c) Final relative density $\rho_i^{\infty} / \rho^{\infty}$ as a function of $r$, for both spontaneous and induced adopters ($i = 0,1$, respectively). (d) Final fraction of spontaneous adopters $\rho_0^\infty$ as a function of $r$. (e) Normalized maximum speed of spreading $\dot{\rho}_m / [p(1 - r)]$, calculated from the derivative of $\rho(t)$. Shaded areas signal the regime of slow contagion $r > r_{\times}$. Curves correspond to $N = 10^4$ and are averaged over $10^3$ realisations.}
\label{Fig:3}
\end{figure}

Considering both extensions, we have a threshold-driven dynamics with three node states: blocked, susceptible and adopter (Fig.\ref{Fig:1}). At the outset, all nodes are susceptible except for a fraction $r$ that remains blocked. At each time step of the simulation, a randomly selected, susceptible node $i$ adopts spontaneously with probability $p$, otherwise it adopts if at least a fraction $\phi$ of its neighbors has already adopted. If $r = 0$ and $p > 0$ all nodes will eventually adopt (Fig.\ref{Fig:3} a), following a kinetics reminiscent of the approach to a unique ground state in a physics system. On the other hand, if we introduce quenched randomness and stochastic perturbations ($r,p > 0$), our model allows various temporal regimes and a transition from rapid to slow spreading dynamics.

Our threshold model can be studied analytically by extending the framework of approximate master equations (AMEs) for monotone binary-state dynamics developed by Gleeson~\cite{gleeson2013binary,gleeson2011high,gleeson2008cascades,porter2014dynamical}, where the transition rate between susceptible and adoption states only depends on the number $m$ of neighbors that have already adopted. We ignore topological correlations by considering a configuration-model network with degree distribution $p_k$ and average degree $z$. We describe a node by the property vector $\kvec = (k, c)$, where $k = 0, 1, \ldots $ is its degree and $c = 0, 1$ its type, i.e. $c = 0$ is the type of the fraction $r$ of blocked nodes, while $c = 1$ is the type of all nodes that may adopt with threshold $\phi$. Moreover, $\Pk$ is a joint distribution giving the probability that a randomly selected node has property vector $\kvec$. Assuming independence between degrees and types, $\Pk = r p_k$ for $c = 0$ and $\Pk = (1 - r) p_k$ for $c = 1$.

The rules of our model are condensed in the probability $\Fkm dt$ that a $\kvec$-node will adopt in a small time interval $dt$, given that $m$ of its neighbors are already adopters, where,
\begin{equation}
\label{eq:thresRule}
\Fkm =
\begin{cases}
p & \text{if} \quad m < k \phi \\
1 & \text{if} \quad m \geq k \phi
\end{cases}, \quad \forall m \; \text{and} \; k > 0,
\end{equation}
with $F_{(k,0),m} = 0$ $\forall k, m$ and $F_{(0,1),0} = p$ (for blocked and isolated nodes, respectively). The dynamics of adoption is well described by an AME for the fraction $\skm(t)$ of $\kvec$-nodes that are susceptible at time $t$ and have $m=0,\ldots,k$ adopting neighbors~\cite{gleeson2013binary,gleeson2011high,porter2014dynamical},
\begin{equation}
\label{eq:AMEsThres}
\dskm = -\Fkm \skm -\bs (k - m) \skm + \bs (k - m + 1) \skmo,
\end{equation}
where,
\begin{equation}
\label{eq:BetaEq}
\bs = \frac{\sumk \Pk \summ (k - m) \Fkm \skm}{\sumk \Pk \summ (k - m) \skm}.
\end{equation}

To reduce the dimensionality of Eq.~(\ref{eq:AMEsThres}) we focus on $\rho(t)$, the fraction of adopters in the network, and $\nu(t)$, the probability that a randomly chosen neighbor of a susceptible node is an adopter. We consider the ansatz $\skm = \Bkm (\nu) e^{-pt}$ for $m < k\phi$ with the binomial distribution $\Bkm(\nu) = \binom{k}{m} \nu^m (1 - \nu)^{k - m}$, leading to the condition $\dot{\nu} = \bs (1 - \nu)$. Then, the AME system is reduced to the pair of ordinary differential equations (see SM),
\begin{subequations}
\label{eq:reducedAMEs}
\begin{align}
\dot{\rho} &= h(\nu, t) - \rho, \\
\dot{\nu} &= g(\nu, t) - \nu,
\end{align}
\end{subequations}
with initial conditions $\rho(0) = \nu(0) = 0$. Here,
\begin{equation}
\label{eq:hFactor}
h(\nu, t) = (1 - r) \Big[ \ft + (1 - \ft) \sum_k p_k \sum_{m \geq k\phi} \Bkm(\nu) \Big],
\end{equation}
and,
\begin{equation}
\label{eq:gFactor}
g(\nu, t) = (1 - r) \Big[ \ft + (1 - \ft) \sum_k \frac{k}{z} p_k \sum_{m \geq k\phi} \Bkom(\nu) \Big],
\end{equation}
with $\ft = 1 - (1 - p) e^{-pt}$. A linear stability analysis of the reduced AME system recovers the cascade condition for $p=0$~(\ref{eq:critical2}) (see SM). Moreover, the fraction of adopters $\rho(t)$ obtained by solving Eq.~(\ref{eq:reducedAMEs}) is in considerable agreement with numerical simulations (Fig.~\ref{Fig:3} a and b). 
Since susceptible nodes adopt spontaneously with rate $p$, their fraction $\rho_0(t)$ in the network is approximated by,
\begin{equation}
\label{eq:innovFrac}
\rho_0 = p \int_0^t (1 - r - \rho) dt.
\end{equation}
where $\rho(t)$ follows Eq.~(\ref{eq:reducedAMEs}) (Fig.~\ref{Fig:3} c and d). We denote its counterpart $\rho_1 = \rho - \rho_0$ as the fraction of {\it induced} adoptions, i.e. vulnerable and stable adopters.

For $p > 0$ the dynamics has a trivial asymptotic state with a final fraction of adopters $\rho^{\infty} = 1 - r$, however, the kinetics of the model depends on the parameters. We first focus on the frequency $F_g$ of global cascades (i.e. adoption reaching at least $20\%$ of susceptible nodes \cite{watts2002simple}) and its behavior in $(\phi, z)$-space for varying $p$ and $r$. For fixed $t$ and $p > 0$, there is a region where global cascades occur (Fig.~\ref{Fig:2} b) that can be compared with the asymptotic cascade regime found for $p = 0$. The boundary of this regime is well approximated by Eq.~(\ref{eq:critical2}) for $p = 0$ and by Eq.~(\ref{eq:reducedAMEs}) for $p \geq 0$. By continuously introducing spontaneous adopters the global cascade regime expands, meaning that macroscopic adoption is eventually possible for systems with any degree and threshold. Even in the absence of a percolating vulnerable component in the network, a growing number of spontaneous adopters induces local cascades that merge due to triggered stable adoptions and finally form a giant component. This behavior is consistent with empirical observations in the online spreading of communication technologies \cite{karsai2015anatomy}.

The kinetics of spreading may change by introducing many blocked nodes. As $r$ (and thus random quenching) increases the adoption process slows down (Fig.~\ref{Fig:3} b). In this dynamics nodes change state in two ways: i) via spontaneous adoption (a slow process for small $p$), or ii) via induced adoption by fulfilling the threshold condition, which may lead to fast cascading behavior. For small $r$ induced adoptions dominate spreading (Fig.~\ref{Fig:3} c) and $\rho$ grows rapidly towards $\rho^{\infty}$. On the other hand, for large $r$ adoption slows down since stable nodes have more blocked neighbors and it is difficult to fulfil their threshold condition. This slow regime is mostly driven by spontaneous adoption, as evidenced by the relatively large asymptotic fraction of spontaneous adopters $\rho_0^{\infty}$ (Fig.~\ref{Fig:3} d).

Taking the ER network as example, a giant component of susceptible nodes can only exist for $r < r* = 1 - 1/z$ \cite{callaway2000network,cohen2000resilience}. Then, a relevant question is whether regimes of fast and slow spreading are separated by a characteristic value $r_\times < r*$. One possibility is to define $r_\times$ as the value that maximizes $\rho_0^{\infty}$ and for which $\rho_0^{\infty} \sim \rho_1^{\infty}$, with $\rho_1^{\infty}$ the final fraction of induced adopters. For $z = 7$, $\phi = 0.2$ and $p = 0.0005$ we have $r_\times \approx 0.7$ and $r* = 0.857$, meaning that slow spreading occurs even in susceptible networks that are not fragmented. The slow regime is further characterized by the lowest possible value in the maximum spreading speed, $\dot{\rho}_m \sim p (1 - r)$, corresponding to the rate of spontaneous adoption at the beginning of the dynamics (Fig.~\ref{Fig:3} e). In other words, the time series $\rho(t)$ has an inflection point for $r < r_{\times}$ and is concave for $r > r_{\times}$.

\begin{figure}
\centering
\includegraphics[width=0.7\linewidth]{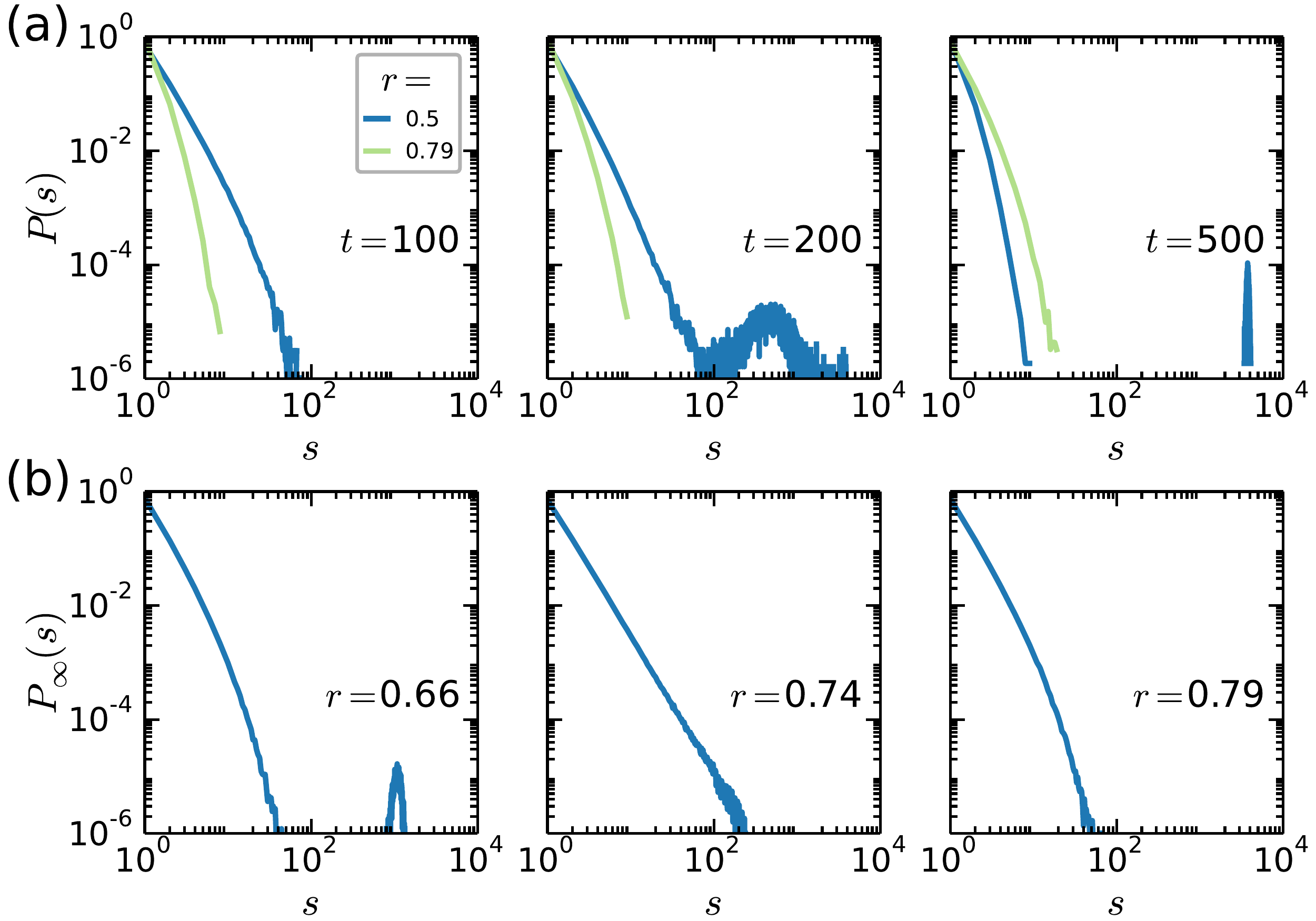}
\caption{(color online). (a) Size distribution $P(s)$ of induced clusters in the regimes of fast ($r < r_{\times}$) and slow ($r > r_{\times}$) spreading, at several stages in the adoption process. In early times, $P(s)$ is unimodal and qualitatively similar in both regimes. As $t$ increases, the distribution becomes bimodal only for $r < r_{\times}$, indicating the presence of global cascades. (b) Asymptotic size distribution $P_{\infty}(s)$ of induced clusters, after $t = 5000$ and for varying $r$. For $r < r_{\times}$ global cascades may still develop and make $P_{\infty}(s)$ bimodal. As $r$ increases, the distribution becomes unimodal and global cascades disappear. Simulations correspond to $z=7$, $\phi=0.2$, $p=0.0005$, $N=10^4$, and are averaged over $10^4$ realisations.}
\label{Fig:4}
\end{figure}

To better understand the kinetics of the crossover between spreading regimes around $r_{\times}$, we finally focus on the size distribution $P(s)$ of induced clusters, i.e. connected components of adopters disregarding spontaneous adopters (Fig.~\ref{Fig:1} b). For early times $P(s)$ includes small induced clusters only, indicating that a larger fraction of spontaneous adopters is crucial for global spreading in the absence of a percolating vulnerable component (Fig.~\ref{Fig:4} a). However, for late times the behavior of $P(s)$ differs between regimes: in the regime of rapid spreading the distribution becomes bimodal due to the appearance of a global cluster of induced adopters, while in the slow regime ($r > r_{\times}$) it remains unimodal until the end of the dynamics. Overall, the crossover between regimes seen globally in the speed of spreading (Fig.~\ref{Fig:3}) is accompanied by an underlying, {\it percolation-type} transition revealed by the asymptotic size distribution $P_{\infty}(s)$ (Fig.~\ref{Fig:4} b). Indeed, in the asymptotic limit $t \to \infty$ and as $r$ increases, this distribution stops being bimodal at $r \approx 0.74 \sim r_{\times}$ for the studied ER case.

The peculiarity of this dynamic percolation transition of induced clusters is that, in contrast to static percolation problems, it is not known {\it a priori} which node will participate in the process, as any unblocked node may become a spontaneous innovator. By analyzing the properties of this transition, we find a critical percolation point at $r_c \sim 0.738$ with the set of exponents $\beta = 1.1$, $\gamma=1.0$, $\tau =2.5$, and $\nu =3.1$, which are rather close to the mean field values~\footnote{For notation and the related theory see D. Stauffer and A. Aharony, {\it{Introduction to Percolation Theory}} (Taylor and Francis, London, 1994); except that the exponent $-(1/\nu)$ is defined here not by the scaling with the linear dimension of the system but with its size.}.

Our aim in this paper has been to provide a general dynamic model of social spreading phenomena that accounts for various kinetics. Our model is designed such that it: (a) is driven by threshold mechanisms capturing the role of social pressure, and (b) concerns temporal aspects of the emergence of global cascades. We generalized Watts' threshold model \cite{watts2002simple} with mechanisms of spontaneous adoption and complete reluctance to adoption, in order to further understand the temporal behavior of spreading phenomena. We have shown that, outside of the cascading regime of the Watts model, there is possibility of global contagion mediated by spontaneous adopters. However, the speed of spreading depends strongly on the density of blocked or immune nodes. For a small fraction $r$ of blocked nodes, few spontaneous adopters enable the formation of large clusters by initiating cascades. For large $r$, spreading slows down as it is dominated by spontaneous adopters and only small cascades are generated. Our intrinsically dynamic model is able to describe various scenarios of real social contagion as well as the crossover between them, and shows a novel percolation transition of induced clusters. This model has not only the potential to explain observational data~\cite{karsai2015anatomy} but, with appropriate fitting, may help identify the character of spreading processes at an early stage, hinting in this way at possible measures to improve adoption performance. Moreover, it is possible that the consideration of blocked nodes will help understand a diversity of spreading phenomena, including related seismic or neural processes.

\vspace{.1in}

RZ acknowledges support from FP7 MULTIPLEX Grant No. 317532, GI from the Academy of Finland, and JK from H2020 FETPROACT-GSS CIMPLEX Grant No. 641191. 

\bibliographystyle{naturemag}
\bibliography{references}

\newpage
\appendix

\begin{center}
{\LARGE Supplemental Information for}\\[0.7cm]
{\Large \textbf{Kinetics of social contagion}}\\[0.5cm]
{\large Z. Ruan, G. I\~niguez, M. Karsai, J. Kert\'esz}\\[2cm]
\end{center}

In this Supplemental Information we present two different schemes of analytical treatment for the dynamical threshold model introduced in the main text. First, in Section \ref{sec:p0} we solve the model exactly for the case of a single innovator seed ($p = 0$), but allowing for a non-zero fraction of blocked nodes to be present in the system ($r \geq 0$ ). We do this by extending the generating function approach provided by Watts \cite{watts2002simple}. Second, in Section \ref{sec:angen} we provide an approximate solution for the general case of multiple innovator seeds introduced periodically in time ($p \geq 0$) and a non-zero fraction of blocked nodes ($r \geq 0$). This solution is based on the approximate master equation (AME) formalism introduced by Gleeson \cite{porter2014dynamical,gleeson2013binary,gleeson2008cascades,gleeson2011high}.

\section{Generating function approach for \texorpdfstring{$p=0$}{p=0}}
\label{sec:p0}

Here we consider the case $p = 0$, $r \geq 0$, i.e. there are no spontaneous adopters in the network (except for a single seed), but a fraction $r$ of blocked nodes hinders the spreading process. Similarly to the Watts case \cite{watts2002simple,singh2013threshold}, the phase diagram can be explored as a static, percolation problem. Our goal is the condition for the existence of a giant vulnerable component \cite{watts2002simple}, which can be found by using the generating function approach \cite{newman2001random}. Suppose $p_k$ is the probability that a randomly chosen node has $k$ connections, and $\rho_k$ is the probability that a node with degree $k$ satisfies the condition $1/k\geq \phi$, where $\phi$ is the threshold of the node. Since a fraction $r$ of nodes is blocked, the probability of a randomly chosen node being unblocked and satisfying the threshold condition is $\rho_{k}(1-r)$. These nodes are {\it vulnerable}, since they may adopt only if at least one of their neighbours has already adopted. Thus the probability of node with degree $k$ being vulnerable is $p_{k}\rho_{k}(1-r)$. The corresponding generating function is,
\begin{eqnarray}
\label{eq:G0}
G_0(x)=\sum_{k}p_k\rho_k(1-r)x^{k}.
\end{eqnarray}

Another generating function we are interested in is $G_1(x)$, which generates the degree distribution of a vulnerable node $b$ that is a random neighbour of a given node. The probability of choosing node $b$ with degree $k$ is $kp_k/z$, thus we have,
\begin{eqnarray}
\label{eq:G1}
G_1(x)=\frac{\sum_{k}kp_k\rho_k(1-r)x^{k-1}}{z}=\frac{G^{'}_0(x)}{z},
\end{eqnarray}
where $z=\sum_{k}kp_k$ is the average degree of the network.
\medskip

To calculate the size distribution of the connected components consisting of vulnerable nodes, we consider two other generating functions,
\begin{eqnarray}
\label{eq:H}
H_0(x)=\sum_{n}q_nx^{n},\\
H_1(x)=\sum_{n}w_nx^{n},
\end{eqnarray}
where $q_n$ is the probability that a randomly chosen node belongs to a vulnerable cluster of size $n$, and $w_n$ is the probability that a random neighbour of a given node belongs to a vulnerable cluster of size $n$.
\medskip

First let us calculate $H_1(x)$. Notice that a random graph below percolation can be regarded as a tree-like structure (since the probability of containing a loop scales as $N^{-1}$ and is negligible for large $N$ \cite{newman2001random}). Under this assumption, a random neighbour of a given node $a$ can be in one of several states: it may not be vulnerable; it may be a vulnerable node with no edges (other than the one connecting to $a$); it may have one edge connected to another component, two edges connected to two components, etc. Then $H_1(x)$ takes the form,
\begin{eqnarray}
\label{eq:H1}
H_1(x)&=&\sum_{n}w_nx^{n}=E(x^n)\nonumber\\
&=&P_1+xG_1(H_1(x)) \nonumber\\
&=&1-G_1(1)+xG_1(H_1(x)),
\end{eqnarray}
where $P_1$ is the probability that a random neighbour of $a$ is not vulnerable, meaning that the node belongs to a vulnerable component with size $n = 0$. According to Eq.~(\ref{eq:G1}), $G_1(1)$ is the probability that a random neighbour of $a$ is vulnerable, thus we have $1 - G_1(1) = P_1$. The term $xG_1(H_1(x))$ comes from the fact that $H_1(x)$ satisfies a self-consistency condition~\cite{newman2001random}. Furthermore, $H_0(x)$ can be calculated in a similar way,
\begin{eqnarray}
\label{eq:H0}
H_0(x)=1-G_0(1)+xG_0(H_1(x)).
\end{eqnarray}
\medskip

\begin{figure}[t]
\centering
\includegraphics[width=0.5\linewidth, trim=0.5cm 0.5cm 0.5cm 0.5cm, clip=true]{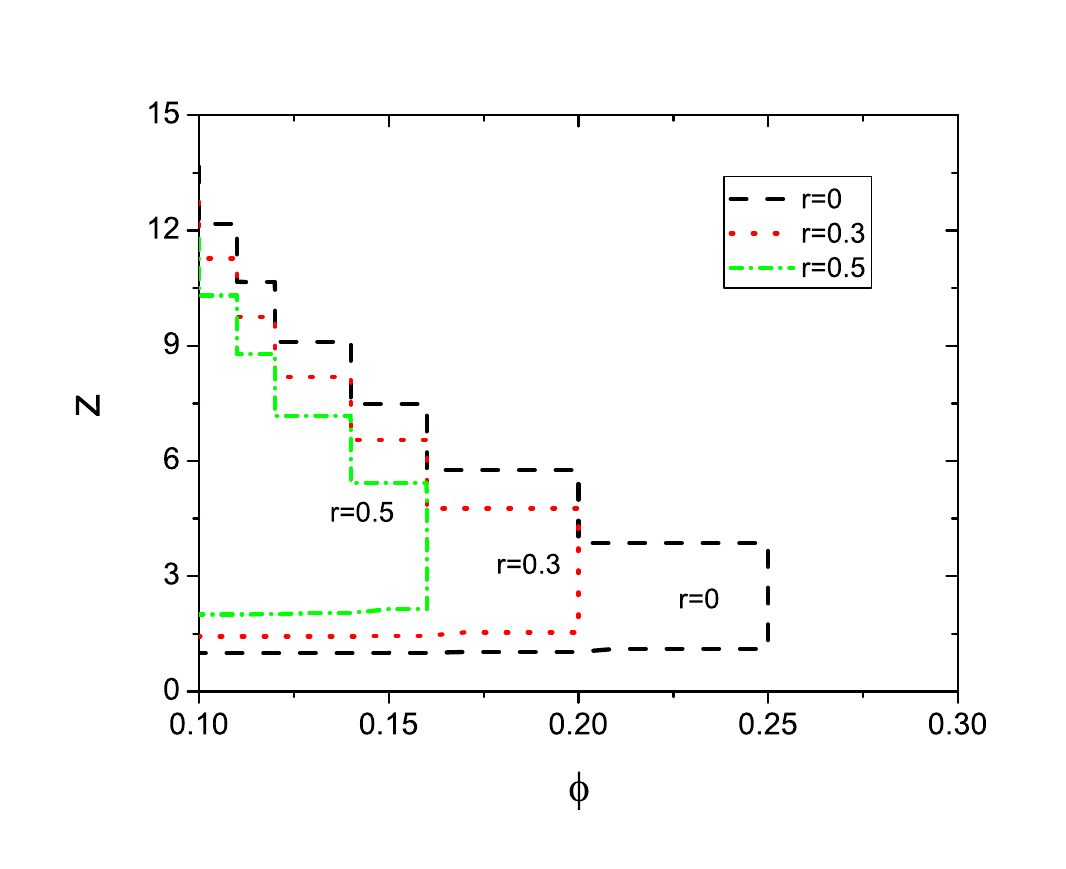}
\caption{Cascade window for different values of $r$. Dashed lines separate two different phases: in the inner region global cascades may happen, but not outside.}
\label{Fig:S1}
\end{figure}

The average vulnerable cluster size is $\langle n \rangle=H^{'}_{0}(1)$. Using Eqs.~(\ref{eq:G1}), (\ref{eq:H1}) and (\ref{eq:H0}) we have,
\begin{eqnarray}
\label{eq:size}
\langle n\rangle=G_0(1)+\frac{(G_0{'}(1))^{2}}{z-G_0^{''}(1)}.
\end{eqnarray}
This expression is similar to the one found for the Watts model \cite{watts2002simple}. However, we have to use Eq.~\ref{eq:G0} for $G_0$. Thus $\langle n\rangle$ diverges for,
\begin{eqnarray}
\label{eq:critical}
G_0^{''}(1)=\sum_{k}k(k-1)(1-r)\rho_kp_k=z.
\end{eqnarray}
When $G_0^{''}(1) < z$ all vulnerable clusters are small since $\langle n \rangle$ is finite, while for $G_0^{''}(1)>z$ there is a giant vulnerable cluster percolating throughout the system.
\medskip

In principle we may solve Eq.~(\ref{eq:critical}) for arbitrary $p_k$ and $\phi$ (or $\rho_k$). We now restrict ourselves to the special case of a Poisson distribution $p_k$ (corresponding to an Erd\H{o}s-R\'{e}nyi random network) and constant $\phi$ (i.e. all nodes have the same threshold). In this case, $p_k=e^{-z}z^{k}/k!$ and $\rho_k$ satisfies,
\begin{eqnarray}
\label{eq:rho}
\rho_k=
\begin{cases}
1, ~~k\leq k_c\\
0, ~~k > k_c
\end{cases},
\end{eqnarray}
where $k_c=\lfloor 1/\phi \rfloor$. Substituting the expressions of $p_k$ and $\rho_k$ into Eq.~(\ref{eq:critical}) we get,
\begin{eqnarray}
\label{eqS:critical2}
(1-r)e^{-z}\sum_{k=2}^{k_c}\frac{z^{k}}{(k-2)!}-z=0
\end{eqnarray}
an expression that can be solved numerically. Notice that there are $3$ parameters here, $z$, $r$ and $\phi$. We may, for example, vary any two of them ($z$ and $\phi$) and solve for the third one ($r$).
\medskip

As it is shown in Fig.\ref{Fig:S1}, the cascade-allowing phase in $(\phi,z)$ space depends on the value of $r$. Each dashed line encloses the region in which the cascade condition is satisfied for a given $r$. As $r$ increases, the cascade window shrinks in both $\phi$- and $z$-axes, as blocked nodes hinder the formation of a giant vulnerable component. The lower boundary of each phase diagram is mainly constrained by the connectivity of the network (since most nodes satisfy the threshold condition because of their low degree) and can be estimated by a mean-field approximation. In average each node has $z$ connections, and among these $z$ neighbors there are $zr$ blocked nodes, meaning that the effective degree of a node is $z-zr$. The condition for the existence of a giant vulnerable cluster is an average effective degree larger than $1$. Thus we have $z=1/(1-r)$, determining the lower boundary of the cascade window. For $r=0.5$ we have $z=2$, which agrees well with calculations shown in Fig.\ref{Fig:S1}.
\medskip

\begin{figure}[t]
\centering
\includegraphics[width=0.8\linewidth]{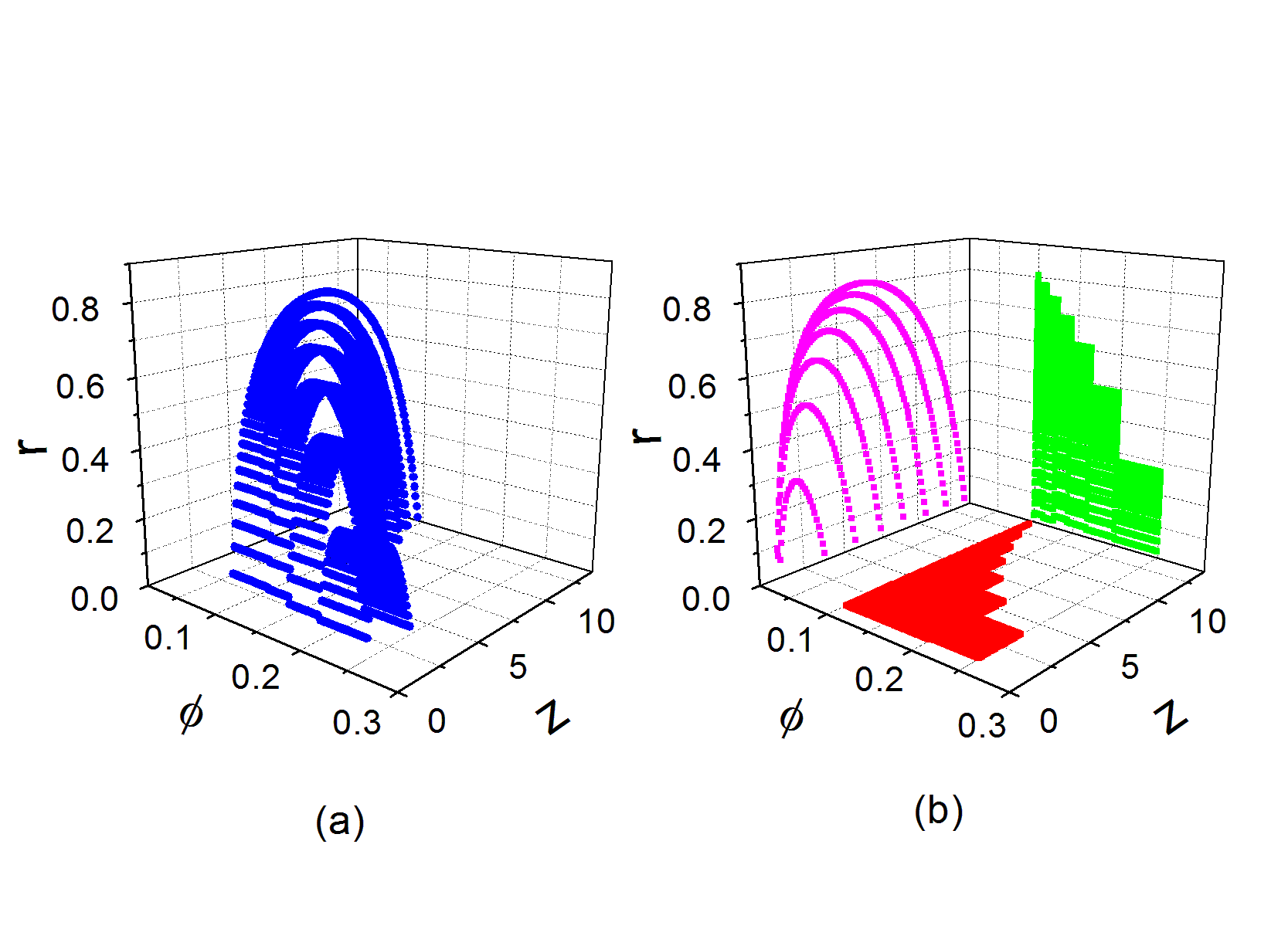}
\caption{(a) Critical $r$ value allowing for the emergence of giant adoption clusters as a function of $\phi$ and $z$. (b) Projections of subplot (a) on the $(\phi,z)$, $(\phi,r)$ and $(z,r)$ planes. }
\label{Fig:S2}
\end{figure}

In Fig.\ref{Fig:S2} (a) we show the critical $r$ value coming from Eq.~(\ref{eq:critical}) as a function of the average degree $z$ and threshold $\phi$ (only positive values are viable solutions). According to Eq.~(\ref{eq:critical}), the calculated critical $r$ values determine a contour surface below which there is a giant vulnerable component; otherwise all vulnerable clusters are small. Fig.\ref{Fig:S2} (b) shows the projections of this surface on the $(\phi,z)$, $(\phi,r)$ and $(z,r)$ planes. Notice that in the projection on the $(z,r)$ plane, $r$ as a function of $z$ first increases and then drops down. The reason is that when the average degree $z$ is very small or very large, the corresponding largest vulnerable cluster is small, i.e., only a few blocked nodes are needed to destroy it.

\section{AME formalism for the general case}
\label{sec:angen}

\subsection{Stochastic binary-state dynamics}
\label{ssec:StochBin}

Here we extend a general rate equation formalism for stochastic binary-state dynamics as developed recently by Gleeson~\cite{porter2014dynamical,gleeson2013binary,gleeson2008cascades,gleeson2011high}. In a stochastic binary-state dynamics, each node in the network can take one of two possible states (susceptible or adopter in the language of innovation adoption) at any point in time, and switching state randomly with probabilities that only depend on the current state of the updating agent and on the states of its neighbors. This general definition includes our threshold model as a special case. Such formalism considers configuration-model networks, that is, an ensemble of networks specified by the degree distribution $p_k$ but otherwise maximally random (i.e. pairs of stubs are connected uniformly at random, so that in the limit $N \to \infty$ of very large network size there are no degree-degree correlations or clustering).
\medskip

For each node to describe all of their relevant properties we introduce a vector $\kvec = (k, c)$, where $k = 0, 1, \ldots k_M$ is the degree of the node and $c = 0, 1$ a dummy variable that labels its `type', i.e. any other property that characterizes the node apart from its degree. In the case of our threshold model, $c = 0$ is the type of the fraction $r$ of blocked (or immune) nodes, while $c = 1$ is the type of all non-blocked nodes that adopt with threshold $\phi$. The maximum value $k_M$ is tuned to approximate the degree distribution with any level of accuracy. Any pair of nodes with identical values of $\kvec$ are considered equivalent in this level of description, forming a node class with the same average dynamics. Moreover, $p_k$ can be generalized to the joint distribution $\Pk$ giving the probability that a randomly selected node has property vector $\kvec$ (i.e. degree $k$ and type $c$). If blocked nodes are chosen randomly among all nodes, like in our model, then $\Pk = r p_k$ for $c = 0$ and $\Pk = (1 - r) p_k$ for $c = 1$.
\medskip

In the language of innovation adoption, the dynamics of a node is determined by the number $m = 0, 1, \ldots k$ of its neighbors that have already adopted when the node is deciding whether to adopt or not. During a small time interval $dt$, a susceptible node (in state 0) adopts with probability $\Fkm dt$, while an adopter (in state 1) becomes susceptible with probability $R_{\kvec, m} dt$. The functions $\Fkm$ and $R_{\kvec, m}$, known as infection and recovery rates, respectively, determine the temporal evolution of the node class $\kvec$. In the particular case of threshold models, a so-called monotone dynamics, $R_{\kvec, m} = 0$ $\forall\, \kvec, m$ (since no adopters become susceptible again). As for $\Fkm$, the rules of spontaneous and threshold adoption imply,
\begin{equation}
\label{eqS:thresRule}
\Fkm =
\begin{cases}
p & \text{if} \quad m < k \phi \\
1 & \text{if} \quad m \geq k \phi
\end{cases}, \quad \forall m \; \text{and} \; k > 0,
\end{equation}
that is, a node adopts the innovation either spontaneously with rate $p$, or with probability 1 if its number of adopting neighbors equals or exceeds the integer threshold $\Phi = \lceil k \phi \rceil$. Blocked nodes ($c = 0$) have an infection rate of $F_{(k,0),m} = 0$ $\forall k, m$, while for isolated nodes ($k = 0$) $F_{(0,1),0} = p$. In other words, blocked nodes never adopt, and isolated nodes can only adopt spontaneously.
\medskip

Let us now turn to the rate equations for our threshold model, called AMEs in the formalism by Gleeson. We denote by $\skm(t)$ the fraction of $\kvec$-class nodes that are susceptible at time $t$ and have $m$ adopting neighbours. Therefore, the fraction of agents with property vector $\kvec$ that are adopters at time $t$ is $\pk(t) = 1 - \sum_{m=0}^k \skm (t)$, and the fraction of adopters in the system is $\rho (t) = \sumk \Pk \pk(t)$. Here, the sum over classes means a sum over all degrees and types, i.e. $\sumk \bullet = \sum_k \sum_c \bullet$. Assuming a monotone dynamics ($R_{\kvec, m} = 0$), the AMEs for $\skm$ can be written as~\cite{porter2014dynamical,gleeson2013binary,gleeson2011high},
\begin{equation}
\label{eqS:AMEsThres}
\frac{d \skm}{dt} = -\Fkm \skm -\bs (k - m) \skm + \bs (k - m + 1) \skmo,
\end{equation}
where $m = 0, \ldots, k$, $s_{\kvec, -1} \equiv 0$, $\Fkm$ follows Eq.~(\ref{eqS:thresRule}), and $\bs(t)$ (the rate at which edges between pairs of susceptible nodes transform to edges between a susceptible agent and an adopter) is given by,
\begin{equation}
\label{eq:rateBs}
\bs(t) = \frac{\sumk \Pk \summ (k - m) \Fkm \skm(t)}{\sumk \Pk \summ (k - m) \skm(t)}.
\end{equation} 
If at time $t = 0$ there is an infinitesimally small seed for the adoption process (i.e. $\rho(0) = 0$), the initial conditions for Eq.~(\ref{eqS:AMEsThres}) are $\skm (0) = \Bkm (0)$, with $\Bkm$ a binomial factor,
\begin{equation}
\label{eq:BinomFac}
\Bkm(\rho) = \binom{k}{m} \rho^m (1 - \rho)^{k - m}.
\end{equation}

The solution $\skm(t)$ of the AME system in Eq.~(\ref{eqS:AMEsThres}) provides a very accurate description of the dynamics of our model, yet its dimension is $(k_M + 1)(k_M + 2)$. Therefore, the number of equations to solve grows quadratically with the maximum degree. Fortunately, the AMEs for our model can be mapped to a reduced-dimension system with a derivation similar to the one used by Gleeson in the case of the Watts threshold model~\cite{watts2002simple,singh2013threshold}.

\subsection{Reduced-dimension AMEs}
\label{ssec:RedDimAMEs}

To reduce the dimension of Eq.~(\ref{eqS:AMEsThres}), we need to consider system-wide quantities that are more aggregated than $\skm$. One of them is the probability that a randomly chosen node is an adopter, $\rho(t) = 1 - \sumk \Pk \summ \skm (t)$, i.e. the fraction of adopters in the network. The other one is the probability that a randomly chosen neighbour of a susceptible node is an adopter, $\nu(t) = \sumk \Pk \summ m \skm(t) / \summ k \skm(t)$.
\medskip

We start by proposing an exact solution for the AME system in terms of the following ansatz,
\begin{equation}
\label{eq:AMEansatz}
\skm(t) = \Bkm [\nu(t)] e^{-p t}
\quad \text{for} \; m < k\phi  \; \text{and} \; c = 1,
\end{equation}
and $s_{(k,0),m} = \Bkm(\nu)$ for $c = 0$, where $\Bkm$ follows Eq.~(\ref{eq:BinomFac}). The meaning of the ansatz in Eq.~(\ref{eq:AMEansatz}) is quite intuitive and considers two processes. First, a susceptible agent with degree $k$ and $m$ adopting neighbours is connected to $m$ adopters with the binomially distributed probability $\Bkm(\nu)$. Second, for $m < k\phi$ a susceptible node does not fulfill the threshold rule and can only adopt spontaneously with probability $e^{-p t}$, since the system is progressively been filled by adopters. Considering these two processes as independent we end up with the product in Eq.~(\ref{eq:AMEansatz}). Finally, since blocked nodes do not adopt and are distributed randomly over the network, $s_{(k,0),m}$ is determined only by a binomial factor.
\medskip

The next step is to insert the ansatz~(\ref{eq:AMEansatz}) into the AME system~(\ref{eqS:AMEsThres}) and derive a set of differential equations for the aggregated quantities $\rho$ and $\nu$. Taking the time derivative $\dskm$ of Eq.~(\ref{eq:AMEansatz}) (i.e. the left-hand side of Eq.~(\ref{eqS:AMEsThres})) we get,
\begin{equation}
\label{eq:ansatzINames1}
\dskm = \left( \left[ \frac{m}{\nu} - \frac{k-m}{1-\nu} \right] \dot{\nu} - p \right) \skm.
\end{equation}
Then, we use the threshold rule~(\ref{eqS:thresRule}) for $m < k\phi$, the ansatz~(\ref{eq:AMEansatz}) and the binomial identity,
\begin{equation}
\label{eq:binomIdent}
\Bkmo(\nu) = \frac{1-\nu}{\nu} \frac{m}{k-m+1} \Bkm(\nu),
\end{equation}
in the right-hand side of Eq.~(\ref{eqS:AMEsThres}) to obtain,
\begin{equation}
\label{eq:ansatzINames2}
-\Fkm \skm -\bs (k - m) \skm + \bs (k - m + 1) \skmo = \left[ -p + \bs \left( m - k + \frac{1-\nu}{\nu}m \right) \right] \skm.
\end{equation}
Equating Eqs.~(\ref{eq:ansatzINames1}) and~(\ref{eq:ansatzINames2}) as in the AME system~(\ref{eqS:AMEsThres}) leads to,
\begin{equation}
\label{eq:condNu}
\dot{\nu} = \bs (1 - \nu),
\end{equation}
a condition on $\nu$ so that the ansatz~(\ref{eq:AMEansatz}) is a solution of Eq.(\ref{eqS:AMEsThres}). This differential equation has the initial condition $\nu(0) = \rho(0) = 0$, obtained by evaluating Eq.~(\ref{eq:AMEansatz}) at $t = 0$ and comparing with the expression $\Bkm (0)$, which corresponds to an infinitesimally small amount of initial adopters randomly distributed among $\kvec$ classes. Furthermore, by assuming a (yet to be determined) function $g(\nu, t)$ such that $\dot{\nu} = g(\nu, t) - \nu$, Eq.~(\ref{eq:condNu}) reduces to,
\begin{equation}
\label{eq:condBeta}
\bs = \frac{g(\nu, t) - \nu}{1 - \nu}.
\end{equation}

Now, we consider the following general result derived by Gleeson in~\cite{gleeson2013binary} (Eqs.~(F6)--(F10) therein),
\begin{equation}
\label{eq:GleesonEq}
\sumk \Pk \summ (k - m) \skm = z (1 - \nu)^2,
\end{equation}
with $z = \sum_k k p_k$ the average degree in the network. Eq.~(\ref{eq:GleesonEq}) is valid for functions $\skm$ and $\nu$ that satisfy Eqs.~(\ref{eqS:AMEsThres}) and (\ref{eq:condNu}) respectively, for any $\Fkm$ and random initial conditions on $\skm$ and $\nu$, and is thus applicable in our case. Our goal here is to use Eq.~(\ref{eq:GleesonEq}) to find an expression for $g(\nu)$ and therefore write the differential equation~(\ref{eq:condNu}) explicitly. Noting that the left-hand side of Eq.~(\ref{eq:GleesonEq}) is the denominator in the definition~(\ref{eq:rateBs}) of $\bs$ and that $F_{(k,0),m} = 0$ (i.e. blocked nodes do not adopt), Eq.~(\ref{eq:rateBs}) gives,
\begin{align}
\label{eq:betaExpl1}
\bs &= \frac{1 - r}{z (1 - \nu)^2} \left[ p \sum_k p_k \summLess (k - m) s_{(k,1),m} + \sum_k p_k \summMore (k - m) s_{(k,1),m} \right] \nonumber\\
&= \frac{1}{z (1 - \nu)^2} \left[ \sumk \Pk \summ (k - m) \skm - r \sum_k p_k \summ (k - m) s_{(k,0),m} \right. \nonumber\\
&\quad \left. - (1 - r) (1 - p) \sum_k p_k \summLess (k - m) s_{(k,1),m} \right],
\end{align}
where we have written $\Pk$ explicitly as $\Pk = r p_k$ for $c = 0$ and $\Pk = (1 - r) p_k$ for $c = 1$, in order to stress the dependence on $r$. Then, we insert the ansatz~(\ref{eq:AMEansatz}) (with its special case $s_{(k,0),m} = \Bkm(\nu)$ for blocked nodes), as well as the identities $(k - m) \Bkm(\nu) = k (1 - \nu) \Bkom(\nu)$ and $\summLess \Bkom(\nu) = 1 - \summMore \Bkom(\nu)$ to obtain,
\begin{equation}
\label{eq:betaExpl2}
\bs = \frac{1}{1 - \nu} \Bigg( (1 - r) \Bigg[ 1 - (1 - p) e^{-p t} + (1 - p) e^{-p t} \sum_k \frac{k}{z} p_k \summMore \Bkom(\nu) \Bigg] - \nu \Bigg).
\end{equation}

A comparison of Eqs.~(\ref{eq:condBeta}) and~(\ref{eq:betaExpl2}) gives us the following expression for $g(\nu, t)$,
\begin{equation}
\label{eqS:gFactor}
g(\nu, t) = (1 - r) \left( \ft + (1 - \ft) \sum_k \frac{k}{z} p_k \summMore \Bkom(\nu) \right),
\end{equation}
where we define $\ft$ as $\ft = 1 - (1 - p) e^{-p t}$. Thus, the AME system~(\ref{eqS:AMEsThres}) gets reduced to the differential equation $\dot{\nu} = g(\nu, t) - \nu$, with $g(\nu, t)$ given explicitly by Eq.~(\ref{eqS:gFactor}).
\medskip

Even though the equation $\dot{\nu} = g(\nu, t) - \nu$ is closed and in this sense equivalent to Eq.~(\ref{eqS:AMEsThres}), we can also derive the corresponding equation for $\rho$, since we are mainly interested in the temporal evolution of the fraction of adopters in the network. From the definition of $\rho$ and Eq.~(\ref{eqS:AMEsThres}) we have,
\begin{align}
\label{eq:rhoDeriv1}
\dot{\rho} = - \sumk \Pk \summ \dskm &= \sumk \Pk \summ \Fkm \skm \nonumber\\
&\quad + \bs \sumk \Pk \summ \big[ (k - m) \skm - (k - m + 1) \skmo \big],
\end{align}
where the second term in the right-hand side telescopes to zero. Then, we use an algebraic manipulation similar to that of Eqs.~(\ref{eq:betaExpl1}) and~(\ref{eq:betaExpl2}) to obtain,
\begin{align}
\label{eq:rhoDeriv2}
& \sumk \Pk \summ \Fkm \skm = (1 - r) \left( p \sum_k p_k \summLess s_{(k,1),m} + \sum_k p_k \summMore s_{(k,1),m} \right) \nonumber\\
&= (1 - r) \left( 1 - (1 - r) (1 - p) \sum_k p_k \summLess s_{(k,1),m} \right) - \rho \nonumber\\
&= (1 - r) \left( \ft + (1 - \ft) \sum_k p_k \summMore \Bkm(\nu) \right) - \rho.
\end{align}
In this way, Eqs.~(\ref{eq:rhoDeriv1}) and~(\ref{eq:rhoDeriv2}) can be rewritten as $\dot{\rho} = h(\nu, t) - \rho$, where,
\begin{equation}
\label{eqS:hFactor}
h(\nu, t) = (1 - r) \left( \ft + (1 - \ft) \sum_k p_k \summMore \Bkm(\nu) \right).
\end{equation}
\medskip

Joining all of these results, the AME system~(\ref{eqS:AMEsThres}) gets reduced to the system of two ordinary differential equations,
\begin{subequations}
\label{eqS:reducedAMEs}
\begin{align}
\dot{\rho} &= h(\nu, t) - \rho, \\
\dot{\nu} &= g(\nu, t) - \nu,
\end{align}
\end{subequations}
with the quantities $g(\nu, t)$ and $h(\nu, t)$ given explicitly by Eqs.~(\ref{eqS:gFactor}) and~(\ref{eqS:hFactor}).
\medskip

The system~(\ref{eqS:reducedAMEs}) can be solved numerically to obtain $\rho(t)$ and thus characterise the temporal evolution of the adoption process. Let us further separate the fraction of adopters as $\rho(t) = \rho_0(t) + \rho_1(t)$, where $\rho_0$ and $\rho_1$ are the fractions of innovators and induced adopters, respectively. Now consider the identity
\begin{equation}
\label{eq:suscepIdent}
1 - \rho = \sumk \Pk \summ \skm = r + (1 - r) \sum_k p_k \summ s_{(k,1),m} = r + \rho_s,
\end{equation} 
where $\rho_s(t)$ is the fraction of non-blocked, susceptible nodes that can eventually adopt, either spontaneously or not. Since such suceptible nodes adopt spontaneously at a rate $p$, the rate equation for innovators is $\dot{\rho}_0 = p \rho_s$. Then, with Eq.~(\ref{eq:suscepIdent}) we obtain,
\begin{equation}
\label{eq:innovRateEq}
\rho_0(t) = p \int_0^t [1 - r - \rho(t)] dt,
\end{equation}
which can be calculated explicitly with the numerical solution of Eq.~(\ref{eqS:reducedAMEs}).

\subsection{Cascade condition for \texorpdfstring{$p=0$}{p=0}}
\label{ssec:p0}

We may also use the AME formalism in the case $p=0$ (no spontaneous adoption) to derive a cascade condition similar to Eq.~(\ref{eq:critical}), as has been done previously for the Watts model \cite{porter2014dynamical}. First we observe that by assuming $p = 0$, $\ft = 0$ and thus we remove the explicit time dependence in Eq.~(\ref{eqS:reducedAMEs}), i.e. $g = g(\nu)$ and $h = h(\nu)$. Then the sum over $m$ in $g(\nu)$ can be rewritten as,
\begin{equation}
\label{eq:gResponse}
g(\nu) = (1 - r) \sum_k \frac{k}{z} p_k \summMore \Bkom(\nu) = (1 - r) \sum_k \frac{k}{z} p_k \sum_{m=0}^{k-1} \Bkom(\nu) f(k,m),
\end{equation}
where,
\begin{equation}
\label{eq:respFunction}
f(k,m) = \begin{cases}
0 & \text{if} \quad m < k \phi \\
1 & \text{if} \quad m \geq k \phi
\end{cases},
\end{equation}
and $f(0,0) = 0$. Eq.~(\ref{eq:respFunction}) is the so-called {\it response function} of the monotone dynamics in our model, i.e. a function that activates when a node with degree $k$ and $m$ adopting neighbours fulfils the threshold condition and therefore may adopt.
\medskip

Let us now perform a linear stability analysis of the reduced AME system~(\ref{eqS:reducedAMEs}) around the equilibrium point $(\rho_*, \nu_*) = (0,0)$, corresponding to a total lack of adoption. If $(\rho_*, \nu_*)$ is unstable, then any small perturbation (like a single node adopting at $t = 0$) can drive the system out of equilibrium and create a global cascade of adoption where $\rho > 0$, that is, a system where a non-vanishing fraction of nodes has adopted in the limit $N \to \infty$. Since the equation $\dot{\nu} = g(\nu) - \nu$ is closed, the stability of Eq.~~(\ref{eqS:reducedAMEs}) is determined by the stability of this equation at $\nu_* = 0$. Then, according to linear stability theory, a local instability exists at $\nu_* = 0$ if,
\begin{equation}
\label{eq:unstabCond}
\frac{d}{d\nu} [ g(\nu) - \nu ] \Big|_{\nu = \nu_*} > 0,
\end{equation}
that is, if $d_{\nu} g(0) > 1$, where $d_{\nu}$ denotes derivative with respect to $\nu$.
\medskip

We can write $d_{\nu}g$ explicitly by inserting the definition of the binomial factor into Eq.~(\ref{eqS:gFactor}),
\begin{equation}
\label{eq:derivGfactor}
\frac{dg}{d\nu} = (1 - r) \sum_k \frac{k}{z} p_k \sum_{m=0}^{k-1} f(k,m) \frac{d\Bkom}{d\nu},
\end{equation}
with
\begin{equation}
\label{eq:derivBinom}
\frac{d\Bkom}{d\nu} = \binom{k-1}{m} \big[ m \nu^{m-1} (1 - \nu)^{k-1-m} - (k-1-m) \nu^m (1 - \nu)^{k-2-m} \big].
\end{equation}
Then we analyse terms in the sum over $m$ at the equilibrium $\nu_* = 0$. For $m = 0$, $d_{\nu} B_{k-1,0} (0) = 1 - k$, but $f(k,0) = 0$ for $\phi > 0$ and thus the term in Eq.~(\ref{eq:derivGfactor}) is zero. For $m = 1$ we have $d_{\nu} B_{k-1,1} (0) = k - 1$. Finally, for $m > 1$ we get $d_{\nu} \Bkom (0) = 0$. Overall, the instability condition~(\ref{eq:unstabCond}) gets reduced to,
\begin{equation}
\label{eq:cascadeCond}
(1 - r) \sum_k \frac{k}{z} (k-1) p_k f(k,1) > 1.
\end{equation}
This condition defines the area in $(\phi,z)$-space where global cascades may develop, for a given value of $r$. By comparing Eqs.~(\ref{eq:rho}) and~(\ref{eq:respFunction}) we see that $f(k,1) = \rho_k$, and so the boundary described by Eq.~(\ref{eq:critical}) is recovered by Eq.(\ref{eq:cascadeCond}). In other words, the cascade regime for $p=0$ is described accurately by both the generating function approach and the AME formalism.

\end{document}